\documentstyle[12pt]{article}
\newcommand{\beq}{\begin{equation}}
\newcommand{\bea}{\begin{eqnarray}}
\newcommand{\eeq}{\end{equation}}
\newcommand{\eea}{ \end{eqnarray}}
\newcommand{\nn}{\nonumber}

\newcommand{\arm}{asymmetric rotor model}
\newcommand{\arma}{asymmetric rotor model }
\newcommand{\psm}{pseudo-SU(3) model}
\newcommand{\psma}{pseudo-SU(3) model }

\newcommand{\la}{\lambda}

\begin{document}
\newfont{\dtt}{cmr12}
\begin{center}
{\Large \bf Effects of Pairing in the Pseudo-SU(3) Model
}
\\[1 cm]
D. Troltenier, C. Bahri, and J.P. Draayer
\\[0.3 cm]
\,{\sl Department of Physics and Astronomy, Louisiana
State University}\\
  {\sl Baton Rouge, LA 70803--4001, U.S.A.}
\\[2 cm]
\end{center}
 \begin{abstract}
An extended version of the \psma which includes both
spin and proton-neutron
degrees of freedom is used to study the influence of the
pairing interaction
on $K$-band mixing, B(E2) values and quadrupole
moments.
Using the \arma as a backdrop, specific consequences of
a many-particle
shell-model based description of these collective
properties are demonstrated
and fundamental limits of the collective model's approach
are investigated.
Finally, the \psm, including representation mixing
induced by pairing, is
used to calculate the energies of $^{140}$Ce and the
results are compared to
experimental data and other theories.

\end{abstract}
\section{Introduction}
Since the discovery of the pseudo-spin symmetry 25
years ago \cite{AH69,HA69}
it  has attracted the attention of numerous physicists and
led to many successful
applications of the theory (see Ref. \cite{TD94} for a
review).
The \psma \cite{RR73} takes full advantage of the
pseudo-spin symmetry and has
been used in the description of a wide spectrum of nuclear
physics phenomena,
ranging from collective excitation spectra \cite{DW84}
and the scissors mode
\cite{CD87}, to identical superdeformed bands
\cite{NT90}, and most recently to
$\beta\beta$-decay \cite{CH94}.

In spite of these successful applications, most calculations
that have used the
pseudo-SU(3) model are very schematic in the sense that:
A) the Hamiltonian has a rather simple structure as it
usually includes only a
long-range quadrupole-quadrupole term plus a residual
interaction designed to
accommodate certain special properties of low-energy
collective spectra, and
B) the configuration space was severely truncated,
usually down to only one
irreducible representation (irrep) of $SU(3)$.
The reason for invoking these simplifications was
technical, in particular, it
was not possible to calculate the matrix elements between
states
of different $SU(3)$ irreps.
Very recently a code was released that lifts this limitation
\cite{BD94}.
It allows:
A) the inclusion of short-range interactions, like pairing
correlations, which
other shell-model theories indicate to be essential, and
B) larger configuration spaces which include many
$SU(3)$ irreps.

As a consequence of these developments, in a recently
published paper
\cite{TD94a} an extended version of the \psm, which
takes the spin degrees
of freedom in a full proton-neutron formalism into
account, was reported.
Specifically, the pairing interaction was expressed in
terms of $SU(3)$
tensor operators and the effects of pairing correlations on
the low-energy
collective spectra and moments of inertia were
demonstrated.

It is the purpose of this contribution to extend these
studies by investigating
the influence of pairing on $K$-band mixing, B(E2)
values, and quadrupole moments.
These studies will be done by comparing the \psma
results to those of the \arm.
The latter is a simple version of the  geometric collective
model which explains
low-energy nuclear properties in terms of rotations and
surface vibrations of
a liquid-drop type nucleus and neglects all single-particle
degrees of freedom.
As a phenomenological model it has been very successful
in the description of
a variety of low-energy properties of even-even nuclei
\cite{TM91}.
By comparing results of the pseudo-SU(3) and the
asymmetric rotor models, it is
possible to identify fundamental limitations of the
collective model approach
on the one hand, while demonstrating the ability of the
\psma to describe collective
nuclear properties correctly on the other.
These schematic studies provide an understanding of the
intrinsic properties of
the \psma that is the basis upon which extended
application of the theory for
describing and predicting nuclear structure phenomena
will rest.
As a first attempt in this direction, the excitation energies
of the semi-magic
nucleus $^{140}$Ce are calculated and compared to
experimental data and the
corresponding results for other theories.

This paper is organized as follows:
In the next section a brief review of the \psma and the
\arma is given, including
a listing of the wavefunctions, Hamiltonians, and
transition operators for
each theory.
In Section 3.1, the importance of $K$, the intrinsic z-axis
angular momentum
projection is discussed, and how states of different $K$
values are coupled in
both the \arma and in the \psma is explored.
In Sections 3.2 the dependence of B(E2) values and
quadrupole moments on the
pairing strength is investigated; specifically, these
measures as revealed by the
\psma are compared to the corresponding \arma results.
In Section 4 the \psma is used to describe the
experimental energy spectrum of
$^{140}$Ce and the results are compared to other
theories.
A summary and conclusion is given in Section 5.

\section{Models and Observables}
An outline of the assumptions and definitions of the wave
functions and
Hamiltonian of the \psma (Subsection 2.1) and \arma
(Subsection 2.2) are
given in this section.
Since the behavior of B(E2) rates and quadrupole
moments are presented in
Section \ref{se:kbq}, definitions of these observables are
given in
Subsection 2.3.

\subsection{Pseudo-SU(3) model}

In a recent publication, an extended version of the \psma
was introduced.
The formulation that was introduced included general
expressions for basis states
and for matrix elements of generic operators between
them \cite{TD94a}.
Therefore, the present discussion is restricted to a brief
summary of the
essential ingredients of the model.
The interested reader can find additional details in Refs.
\cite{DW84,CD87}.

The extended version of the \psma introduced in Ref.
\cite{TD94a} and
which is employed here, explicitly includes the spin
degree of freedom and
the Pauli Exclusion Principle in a full proton-neutron
fermion formalism.
Specifically, the \psma is a $0\hbar \omega$ theory;
vertical couplings to higher
shells are only allowed in an extension of the \psma called
the pseudo-symplectic
model \cite{TD94}.
In this contribution the protons $(\pi)$ and neutrons
($\nu$) occupy the real-space
shells $N_\pi = 4$ and $N_\nu = 5$ or, equivalently, the
pseudo-space shells
$\tilde{N}_\pi = 3$ and $\tilde{N}_\nu = 4$ which is
characteristic of rare-earth
nuclei.

The wavefunctions of the \psma are classified by Casimir
invariants of the group
chain (see Ref. \cite{TD94a})
\bea
&&\nn\\
U(2 \Omega_\pi) \,  \times \, U(2 \Omega_\nu)
&\supset&
\left[
U(\Omega_\pi) \times  U(2)^{(\pi)}
\right]
\, \times \,
\left[
U(\Omega_\nu) \times  U(2)^{(\nu)}
\right]
\nn \\
&&\nn\\
&  \supset&
\left[
SU(3)^{(\pi)} \times  SU(2)^{(\pi)}
\right]
\, \times \,
\left[
SU(3)^{(\nu)} \times  SU(2)^{(\nu)}
\right]
\nn \\
&&\nn\\
&  \supset&
\left[
SU(3)^{(\pi)} \times  SU(3)^{(\nu)}
\right]
\, \times \,
\left[
SU(2)^{(\pi)} \times  SU(2)^{(\nu)}
\right]
\nn \\
&&\nn\\
& \supset &
SU(3) \times  SU(2)_S
\supset
 SO(3)_L \times  SU(2)_S
\supset
SU(2)_J
\, \\
&&\nn
\eea
where $2\Omega_\sigma$ denotes the total number of
single-particle levels in
the $N_\sigma$ shell for protons ($\sigma=\pi$) and
neutrons ($\sigma = \nu$),
respectively.
The irreducible representations (irreps) of
$U(2\Omega_\sigma) $, labeled by
$\left[ 1^{m_\sigma} \right]$, where $m_\sigma$
denotes the number of
$\sigma$-type particles, are the antisymmetrized many-
particle proton or neutron
wavefunctions which separate into spatial
($\rightarrow U(\Omega_\sigma)$, irrep label $\left[
f_\sigma \right]$ ) and spin
($\rightarrow U(2)^{(\sigma)}$,  irrep label $\left[
\bar{f}_\sigma \right]$)
degrees of freedom.
All calculations presented in this contribution are
restricted to the most symmetric
and energetically lowest irrep of $U(\Omega_\sigma)$,
which means that couplings to
$(S \ne 0)$ modes are excluded.
This seems reasonable since those spin-flip excitations are
significantly higher
in energy than the low-energy excitations which are of
primary interest in this
contribution. All $SU(3)$ irreps
$(\lambda_\sigma, \, \mu_\sigma)$ that are contained in a
given
$\left[ f_\sigma \right]$ are determined from the reduction
$U(\Omega_\sigma) \supset SU(3)^{(\sigma)}$ where
possible multiple
occurrences of the same $(\lambda_\sigma, \,
\mu_\sigma)$ are labeled by the integer
$\alpha_\sigma$.
The reduction
$\left[ SU(3)^{\left( \pi \right)} \times SU(3)^{\left( \nu
\right)} \right] \rightarrow
SU(3)$ leads
to the $SU(3)$ irreps of the total wavefunction, labeled
by $(\lambda, \, \mu) $,
which are determined by taking all possible products
$\left\{ (\lambda_\pi, \, \mu_\pi) \times (\lambda_\nu, \,
\mu_\nu) \right\}\,
\rightarrow \, \rho \, (\lambda, \, \mu) $ into account,
where the running
index $\rho = 1, \ldots,\,  \rho_{\mbox{\small max}} $
numbers possible multiplicities.
The orbital angular momentum values $L$ which are
contained in a fixed
$(\lambda,\, \mu)$-irrep are determined through the
reduction
$SU(3) \supset SO(3)_L$ and multiple occurrences are
numbered by the index $\kappa$.
The total spin $S$ is derived from the SU(2) product
$S_\pi \times S_\nu \rightarrow S$ where proton and
neutron spins
$S^\pi$ and $S^\nu$ label the irreps of $SU(2)^\pi$ and
$SU(2)^\nu$ respectively.
Finally the total angular momentum $J$ is obtained
through the SU(2) coupling
$L\times S \rightarrow J$ as symbolized by the group
reduction
$\left[ SO(3)_L \times SU(2)_S \right] \rightarrow
SU(2)_J$.

The eigenvalues of the Casimir invariants provided by
this group reduction scheme,
augmented with the necessary multiplicity labels, allows
for a unique classification
of the \psma wavefunctions,
\begin{eqnarray}
\lefteqn{
| \left\{ m_\pi [f_\pi] \alpha_\pi \, (\la_\pi,\,\mu_\pi) \, ,
          m_\nu [f_\nu] \alpha_\nu \, (\la_\nu,\,\mu_\nu) \,
\right\}
          \rho \, (\la,\,\mu) \kappa L \,
         \left\{S_\pi , S_\nu\right\} \, S \, ; JM \rangle
        }
\nn\\
&&\nn\\
&=&\sum_{\{ -\} }
\langle \,    (\la_\pi,\,\mu_\pi) \,  \kappa_\pi L_\pi
M_{L_\pi}\, ;
              (\la_\nu,\,\mu_\nu) \,  \kappa_\nu L_\nu
M_{L_\nu}\,
              | \,
              (\la,\,\mu) \,  \kappa  L M_L  \,
              \rangle_\rho
\nn \\
&&\nn\\
&&\times \, \langle \, S_\pi M_{S_\pi}, \,
                                                              S_\nu
M_{S_\nu}\,| \,
                                                              S  M_S
      \rangle  \, \times \,
\langle \, L M_L, \,
                  S M_S \,| \,
                   JM
\rangle  \, \nn\\
&&\nn\\
&& \times \,
         |m_\pi [f_\pi]   \alpha _\pi \,  (\la_\pi,\,\mu_\pi) \,
         \kappa_\pi L_\pi M_{L_\pi}\,  S_\pi M_{S_\pi}
         \rangle
\,\times \,
         |m_\nu [f_\nu]   \alpha _\nu \,  (\la_\nu,\,\mu_\nu) \,
         \kappa_\nu L_\nu M_{L_\nu}\,  S_\nu M_{S_\nu}
         \rangle
\nn\\
&&
\label{eq:psmwf}
\end{eqnarray}
with the abbreviation
$\{ - \}  = \{  M_{S_\pi}, \,  M_{S_\nu}, \,   M_{S}, \,
                      M_{L_\pi}, \,  M_{L_\nu}, \,   M_{L},  \,
                       \kappa_\pi, \, \kappa_\nu, \, L_\pi, \, L_\nu
                 \} \, .
$
In this result
$\langle \, L_1 M_{L_1}, \,
            L_2 M_{L_2}\,| \,
            L M_L
\rangle  $
denotes SU(2) Clebsch-Gordan coefficients and the
$SU(3)$ coupling coefficient
$\langle \ldots;\, \ldots | \dots \rangle $
is defined in Ref. \cite{AD73}.
\subsubsection{Hamiltonian and Transition Operators}
The \psma Hamiltonian, $H_{PSU(3)}$, used for both
the schematic calculations in
Section \ref{se:kbq} and a comparison to experimental
data in Section \ref{se:fit},
is a many-particle extension of the Nilsson-model
Hamiltonian,
\beq
H_{PSU(3)} =
H_0
- \frac{\chi}{2} \,\,  Q^a \cdot Q^a
- G_\pi H^\pi_P
- G_\nu H_P^\nu
+ a K_J^2 + bJ^2
+ D_\pi \sum_{i_\pi}l_{i_\pi}^2
+ D_\nu \sum_{i_\nu}l_{i_\nu}^2
\label{eq:psh}
\eeq
where $Q^a$ denotes the algebraic quadrupole operator
and $\chi$ the
quadrupole-quadrupole interaction strength parameter.
($H_0$ denotes the spherical harmonic oscillator which,
however, is trivial since
the configuration space of the pseudo-SU(3) model is
restricted to a single shell
in both proton and neutron spaces.
Since only {\it excitation} energies are of interest, the
$H_0$ contribution, which
is equal for every configuration, is of no physical
importance.)
$H^\pi_P$ ($H^\nu_P$) stands for the pairing interaction
for protons (neutrons)
which is multiplied by the pairing strength parameter
$G_\pi$ $(G_\nu)$.
The term $K_J^2$ is introduced to accommodate the
$K$-band splitting observed in
the low-energy spectra of heavy deformed nuclei and
$J^2$ is the square of the total
angular momentum.
These five terms constitute the Hamiltonian for the
schematic calculations in Section
\ref{se:kbq} and have already been discussed to some
extend in previous publications
so the reader is referred to Refs.
\cite{TD94,ND90,ND92} for additional details.
The parameter set is identical to the one used in Ref.
\cite{TD94a}, namely,
$\chi= 4.32$ keV, $a = 202$ keV, and $b=9.26$ keV,
$D_\pi = D_\nu = 0$
and is compatible with sets obtained from best-fit
calculations for rare earth nuclei
\cite{TE95}.

In the Nilsson Hamiltonian the squared single-particle
angular momentum operator,
$l^2$, is of great importance since it effectively flattens
the harmonic oscillator
potential for states of higher angular momentum and in so
doing mimics a radial
potential shape of the Woods-Saxon type.
It is therefore necessary to add analogous terms to
$H_{PSU(3)}$ if the \psma is
used for the description and prediction of experimental
data.
This is why the Hamiltonian of Ref. \cite{TD94a} has
been extended to include
the terms $ D_\pi \sum_{i_\pi}l_{i_\pi}^2$  and $D_\nu
\sum_{i_\pi}l_{i_\pi}^2$
for protons and neutrons, respectively.
The particular choice of parameters used for the \psma
application to $^{140}$Ce
is discussed in Section \ref{se:fit}.

An expression for the electric quadrupole transition
operator and the magnetic
dipole operator for the \psma is given in Ref.
\cite{CD87}.
The result for the real-space electric quadrupole transition
operator is
\beq
T_{2\mu}(E2) \equiv
b_0^2
\left(
e_\pi \sum_{i\epsilon\pi}
r_\pi^2(i) Y_{2\mu}(\vartheta_i^\pi, \varphi_i^\pi)
+
 e_\nu \sum_{i\epsilon \nu}
r_\nu^2(i) Y_{2\mu}(\vartheta_i^\nu, \varphi_i^\nu)
\right)
\label{eq:te2}
\eeq
where $b_0 = A^\frac{1}{3}$fm is the harmonic
oscillator size parameter,
$e_\sigma$ is the effective nuclear charge for protons
($\sigma=\pi$)
and neutrons ($\sigma=\nu$), respectively (see Section
\ref{se:fit}),
the spherical single-particle coordinates of the valence
nucleons are denoted
by $r_\sigma(i),\vartheta_i^\sigma$, and
$\varphi_i^\sigma$, and $Y_{2\mu}$ stands
for a spherical harmonic \cite{VM88}.
Since all calculations of the \psma are performed in the
pseudo-space, this
real-space operator must be expressed in terms of pseudo-
space quantities.
The explicit procedure and the results of this real-space
$\rightarrow$
pseudo-space transformation are given in Ref.
\cite{CD87}.

\subsection{Asymmetric Rotor Model}
For the sake of completeness, the most important
definitions of the \arma
are given next.
The reader is referred to Refs.
\cite{DF58,DC60,Da65,AB68,EG87} for a more
detailed account.
The \arma Hamiltonian $H_{ARM}$ is defined as
\bea
H_{ARM} &=& \sum_{k=1}^3 \frac{I_k^2}{{\cal
J}_k} \nn\\
&=& \left( I^2 - I_3^2 \right) /
    \left( \frac{1}{4{\cal J}_1} + \frac{1}{4{\cal J}_2}
\right) +
    \frac{I_3^2}{2{\cal J}_3} \nn\\
&&  + \left( I_+^2 + I_-^2 \right) /
      \left( \frac{1}{8{\cal J}_1} - \frac{1}{8{\cal J}_2}
\right)\, ,
\label{eq:harm}
\eea
where $I^2_k$ denotes the intrinsic $k$-component of
the total angular momentum
$I$ and $I_+$ ($I_-$) the corresponding raising
(lowering) operators.
The hydrodynamic moment of inertia ${\cal J}_k \equiv 4
B \beta^2 \sin (3\gamma
- \frac{2 k \pi}{3} )$ depends on the mass parameter $B$
and the quadrupole
deformation variables $(\beta, \gamma)$ (with $0^o \le
\gamma \le 60^o$,
$\beta > 0$, see Ref. \cite{EG87}) all of which are
treated as adjustable model
parameters.
The eigenfunctions of $H_{ARM}$,
$\Psi^{ARM}(\vartheta_i)$, are determined by
diagonalization within an orthonormal basis of
symmetrized wave functions
\bea
|IMK> &\equiv &
\sqrt{\frac{2I + 1 }{16\pi^2 (1 + \delta_{K0}) } }
\left( D_{M K}^{I*} (\vartheta_i) + (-1)^I
       D_{M-K}^{I*} (\vartheta_i)
\right)
\eea
where the Wigner functions
$ D_{MK}^{I*} (\vartheta_i) $ \cite{Ro57} depend on
the three Euler angles
$ (\vartheta_1, \vartheta_2, \vartheta_3) $ and $M$ $(K)$
denotes the
laboratory (intrinsic) angular momentum z-axis
projection.
For a fixed angular momentum $I$ the basis
dimensionality is given by the
number  of possible $K$ values:
$K=0, 2, 4, \ldots, I$ if $I$ even and
$K=2, 4, \ldots, I-1 $ if $I$ odd.
\\[1 cm]
{\bf Figure \ref{fig:armen}}
\\[1 cm]
Finally, recall that the \arma eigenenergies have a
characteristic $\gamma$
dependence (see Fig. \ref{fig:armen})
which is symmetric around $\gamma = 30^o$.
(The other two model parameters $\beta$ and $B$ induce
only an overall scaling
of the eigenenergies.)

The quadrupole operator $Q^{ARM}_{2\mu}$ of the
ARM is given in lowest order by
\bea
Q^{ARM}_{2\mu} = A \beta
\left(
D_{\mu 0}^{2*} (\vartheta_i) \cos\gamma +
\left(
 D_{\mu 2}^{2*} (\vartheta_i) + D_{\mu -2}^{2*}
(\vartheta_i)
\right)
\frac{ \sin \gamma}{\sqrt{2}}.
\right)
\eea
In this expression the normalization coefficient $A = 3Z
R_o^2 /4 \pi$ depends
upon the charge $Z$ and the nuclear radius $R_o$.
\subsection{Physical Observables}
For the sake of completeness, and before continuing with
a more detailed consideration
of the pseudo-SU(3) wave functions, the definitions of
some physically important
quantities will be given.
Reduced matrix elements of the SO(3) tensor operator,
$Q_{\lambda \mu}$, between states
with initial (final) angular momentum and projection
$J_i$ and $M_i$ ($J_f$ and $M_f$)
are defined by \cite{Ro57}
\[
\langle \gamma_f J_f || Q_\lambda || \gamma_i J_i \rangle
\langle J_i M_i, \lambda \mu \, | J_f M_f \rangle \equiv
\langle \gamma_f J_f M_f   | Q_{\lambda \mu} | \gamma_i
J_i M_i \rangle
\]
where $\gamma_i$ and $\gamma_f$ represent additional
quantum numbers that are required
to uniquely define the initial and final state, and  $\langle
\gamma_f J_f || Q_\lambda
|| \gamma_i J_i \rangle$ stands for the reduced matrix
element.
(The use of $J$ for the total angular momentum is
customary in microscopic work while
$I$ is normally used in macroscopic theories.
Here and in what follows, $J$ and $I$ will
be used interchangeably for the total angular momentum.)
The definition of the reduced transition probability for
electric quadrupole radiation
\cite{EG87} is then given as
\beq
B(E2; \, \gamma_i J_i  \, \rightarrow \, \gamma_f J_f )
\equiv
\frac{2J_f+1}{2J_i+1}
\langle \gamma_f J_f \,  || T_{2}(E2) || \,  \gamma_i J_i
\rangle^2 \,
\nn
\eeq
and the
definition of the electric quadrupole moment is
\beq
Q( \gamma \, J )
\equiv
\sqrt{\frac{16 \pi}{5}}
\sqrt{ \frac{J (2 J-1)}{(J+1)(2J+3)} }
\langle \gamma \, J \,  || T_{2}(E2) || \, \gamma \, J \rangle
\, .
\nn
\eeq
\section{Pairing and Observables \label{se:kbq}}
The consequences of the pairing interaction within the
\psma is explored in this
section by considering its effect on $K$-band mixing and
on B(E2) values and
quadrupole moments.
The $K$-band mixing is considered first because the
information provided by this
effect is easier to realize and simpler to understand than
the corresponding B(E2)
and quadrupole moment results.
\subsection{$K$-band Mixing}
The quantum number $K_J$ is the projection of the total
angular momentum $J$
on the system's intrinsic z-axis.
Since only $S=0$ states are considered in this
contribution, $J \equiv L$ and
$K_J \equiv K_L$ and thus the symbol $K$ can and will
be used for both.
The $K$ label is important in the classification of low-
energy collective bands
of prolate or near-prolate even-even nuclei.
Specifically, the ground-state band of these nuclei is
associated with $K=0$,
the so-called one-phonon $\gamma$ band with $K=2$,
likewise the two-phonon
$\gamma$-band with $K=4$
or $K=0$,
and so on.\footnote{Recently the $K$=4 band received
some attention through a systematic measurement of
$\gamma$-vibrational
anharmonicities in the rare-earth region
\cite{BJ91,WA94}}

The quantum number $K$ is exactly conserved for
prolate nuclei only; couplings
between bands with different $K$ values are connected
with deviations from axial
symmetry, that is, with $\gamma \ne 0$.
 From the point of the view of the Geometric Collective
Model \cite{TM91,TM92},
the $K$-band mixing can be understood within the
framework of the very simple
\arma picture:
The Hamiltonian in Eq. \ref{eq:harm} is diagonal so long
as $\gamma = 0$, since
in this case ${\cal J}_1={\cal J}_2$ and $K$ is trivially
conserved.
Increasing $\gamma$ (introducing triaxiality) generates
non-vanishing matrix
elements between basis states with $\Delta K = \pm 2$.
For example, the $K = 0$ eigenstate of the symmetric
rotor gains non-vanishing
contributions from the $K = 2$ basis function, and so on.
This $\beta$-independent phenomenon is illustrated
quantitatively in Fig. \ref{fig:armk}
which depicts the expectation value of $K^2$ for the even
$J$ \arma yrast states,
$\langle ARM | K^2| ARM \rangle_y$, as a function of
$\gamma$.
\\[1 cm]
{\bf Figure \ref{fig:armk}}
\\[1 cm]
The results show that $\langle ARM | K^2| ARM
\rangle_y$ increases with both higher
angular momentum and larger $\gamma$ values:
For a fixed value of the $\gamma$ deformation, $\langle
ARM | K^2| ARM \rangle_y$
increases with increasing angular momentum because the
higher the angular momentum,
the higher the $K$ values ($K \le I$) of basis functions
$|IMK>$ that can contribute
to the eigenfunction.
And for a fixed value of the angular momentum, $\langle
ARM | K^2| ARM \rangle_y$
increases with increasing $\gamma$-value because the
non-diagonal term in $H_{ARM}$
becomes more and more dominant.
On the far right-hand-side of Fig. \ref{fig:armk}, the
expectation value of $K^2$ is
indicated for the case of complete mixing, that is, if all the
diagonalization
coefficients have the same $\frac{1}{\sqrt{d}}$
magnitude and thus
$\langle ARM | K^2| ARM \rangle_y = \frac{1}{d}
\sum_K K^2$, where $d$ denotes
the dimensionality.
At $\gamma=60^o$ these estimates are about $25$
percent below the results of the
diagonalization, indicating that the yrast states favor basis
state with higher
$K$ values than that of a uniform distribution.

It is now interesting to use these simple results as a
backdrop for a study of pairing
induced $K$-band mixing in the \psm.
To accomplish this it is necessary to have a shell-model
expression for the $\hat{K}^2$
operator \cite{ND90,ND92,NB94}.
For completeness the argument that leads to such a result
is repeated here.
The basic idea follows from a frame-independent shell-
model expression for the
\arma Hamiltonian (see Eq. \ref{eq:harm}).
To achieve this one can introduce two SO(3) scalars
\bea
&&\nn\\
X_3^a & \equiv & \sum_{i,j} L_i Q^a_{ij} L_j =
\sum_{i} \lambda_i I_i^2\, , \nn \\
&&\nn\\
X_4^a & \equiv & \sum_{i,j, k } L_i Q^a_{ij} Q^a_{jk}
L_k = \sum_{i} \lambda_i^2 I_i^2 \nn \\
&&\nn
\eea
where $L_i$ ($I_i$) and $Q^a_{ij}$ $(\lambda_i)$
denote, respectively, projections
of the total angular momentum and the algebraic
quadrupole tensor in either the
laboratory or the intrinsic body-fixed principal axis
system \cite{ND90}.
(The sums in these expression insure rotational invariance
and thereby the
frame-independent character of the operators.)
Adding the total angular momentum $L^2 = I^2 =
\sum_{i} I_i^2 $ to $X^a_3$ and $X^a_4$
means a mapping onto the \arma Hamiltonian can be
established:
\[
\frac{I_1^2}{2 {\cal J}_1}
+ \frac{I_2^2}{2 {\cal J}_2} + \frac{I_3^2}{2 {\cal
J}_3} \equiv
a L^2 + b X_3^a + c X_4^a, \,
\]
where the coefficients $a$, $b$, and $c$ are real
numbers.
The geometrical expression for the $\hat{K}^2  = I_3^2
$ operator emerges trivially
from this expression by setting $1/2 {\cal J}_1 \equiv 1/2
{\cal J}_2 \equiv 0$ and
$1/2 {\cal J}_3 \equiv 1$.
Taking advantage of this observation, it is straightforward
to derive a shell-model
expression for $\hat{K}^2$ \cite{ND90}
\beq
\hat{K}^2 = \frac{\lambda_1 \lambda_2 L^2 +
\lambda_3 X_3^a +
           X_4^a}{2\lambda_3^2 + \lambda_1\lambda_2}
\label{eq:k**2}
\eeq
where the principle axis quadrupole components can be
shown to be given
by \cite{ND90}
\[
\lambda_1 = \frac{1}{3}(-\lambda + \mu)\, , \, \,
\lambda_2 = \frac{1}{3}(-\lambda - 2\mu - 3)\, , \, \,
\lambda_3 = \frac{1}{3}(2\lambda + \mu + 3) \, .
\]
The matrix elements of $\hat{K}^2 $ in the extended
version of the \psma
which is used in this contribution can be easily derived
using the generic
formulas in \cite{ND92}.

To investigate the influence of the pairing force strength
on $K$-band
splitting in the \psm, consider a configurations with two
protons and two
neutrons ($2\pi, \, 2\nu$) in the real $[$pseudo$]$ space
shells $(N_\pi
= 4, N_\nu=5)$ $[(\tilde{N}_\pi = 3,
\tilde{N}_\nu=4)]$.
This configuration exhibits a rich structure and its
dimensionalities are
small enough that basis space truncation measures need
not be invoked,
that is, all possible couplings of proton and neutron
$SU(3)$ irreps can be
taken into account for all angular momentum values.
\\[1 cm]
{\bf Figure \ref{fig:2p2nk**2}}
\\[1 cm]
Figure \ref{fig:2p2nk**2} shows the expectation value of
$\hat{K}^2$ for
even angular momentum yrast states of the $(2\pi, \,
2\nu)$ configuration,
$\langle 2\pi 2\nu\, |\,  K^2 \, |\, 2\pi  2\nu\rangle_y$, as a
function of
the pairing strength parameter $G$, where for simplicity
$G_\pi$ and $G_\nu$
were set equal ($G_\pi = G_\nu \equiv G$) in the
calculations.
Before entering into a detailed discussion of these results,
note that only
the pairing terms in Eq. \ref{eq:psh} couple states
belonging to different $SU(3)$
irreps ($\lambda, \mu$).
If $G=0$ the quadrupole-quadrupole interaction
$Q^a\cdot Q^a\sim 4C_2 - 3L^2$
dominates and forces states of the $SU(3)$ irrep with
largest second order
Casimir invariant $C_2$ (Eq. \ref{eq:c2c3}) to lie
energetically lowest.
For the $(2\pi, \, 2\nu)$ configuration this leading irrep is
$(\lambda, \mu) = (14,0)$ which contains $K=0$ states
only and is the reason
why the results show $\langle 2\pi 2\nu | \hat{K}^2 | 2\pi
2\nu\rangle_y = 0$
for all value of the angular momentum (Fig.
\ref{fig:2p2nk**2}).\footnote{The
selection rule for $K$ in a fixed $(\lambda, \mu)$-irrep is
given by $K =$
min$(\la, \mu)$, min$(\la, \mu)$ -2, min$(\la, \mu)-4 ,
\ldots , 0$
or $1$ \cite{CL93}.}

In a recent paper (see Figures 3-5 in \cite{TD94a}) the
intensity distribution
of the yrast eigenstates of $H_{PSU(3)}$ was shown to
spread over more and more
basis states with increasing pairing strength.
As a consequence, the $(2\pi, \, 2\nu)$ yrast states have
contributions from
$(\lambda\ne 0,\mu \ne 0)$ $SU(3)$ irreps which contain
$K \ne 0$ configurations.
This admixture shows up (Fig. \ref{fig:2p2nk**2}) as an
increase in the
$\langle 2\pi 2\nu |\hat{K}^2 | 2\pi 2\nu\rangle_y$ value
with increasing
pairing strength $G$.
(The $L=10, 12, 14 \hbar$ states are unaffected since
they are unique.)

Another mechanism that acts in the same direction is the
coupling to higher
$K$ states within one ($\lambda\ne 0, \, \mu\ne 0)$ irrep
through the $X_3^a$
and $X_4^a$ operators contained in $\hat{K}^2$ (Eq.
\ref{eq:k**2}).
This effect increases the expectation value of
$\hat{K}^2$ for states of higher
angular momentum as long as there are states with higher
$K$ values available.

For fixed $G$ the $\langle 2\pi 2\nu |\hat{K}^2| 2\pi
2\nu\rangle_y$ value
increases up to $J=6\hbar$ and goes down for higher
angular momenta due to
the decreasing availability of higher $K$ basis states.
This mechanism counteracts the two aforementioned
effects and is illustrated
in Fig. \ref{fig:k-av} where the average $K$ value,
$\bar{K}$, for a fixed even
angular momentum value is given for the $(2\pi, \, 2\nu)$
configuration.
\\[1 cm]
{\bf Figure \ref{fig:k-av}}
\\[1 cm]
Note that the $(2\pi, \, 2\nu)$ distribution has its
maximum at $J=4$ and
clearly indicates the decreasing availability of higher $K$
basis states with
increasing $J$.
\\[1 cm]
{\bf Figures \ref{fig:k-av gen}a) and \ref{fig:k-av gen}b)}
\\[1 cm]
General results of this type are shown in Figs.
\ref{fig:k-av gen} where
$\bar{K}$ values for identical particle configurations in
the $\tilde{N} = 3$
(Fig. \ref{fig:k-av gen}a) and $\tilde{N} = 4$ (Fig.
\ref{fig:k-av gen}b) shells
are given.
Only results for particle configurations are given since the
$\bar{K}$ values
for holes are identical to those for particles if the hole and
particle numbers
are equal.
The latter is the reason why within a single shell the
maximum $\bar{K}$ value,
$\bar{K}_{max}$, increases until the shell is half full
and decreases in a
particle-hole symmetric fashion when the shell is more
than half full.
By comparing results for the $\tilde{N} = 3$ and
$\tilde{N} = 4$ shells one
finds that the $\bar{K}$ distributions are shifted towards
higher $K$ values
with increasing shell number and that as a general rule the
maximum average
$\bar{K}$ value for each configuration is found at
roughly one third of the
maximum possible angular momentum.
This feature is a consequence of the selection rule on $K$
values in a fixed
$SU(3)$ irrep, which in turn follows from the Pauli
Exclusion Principle
(see Ref. \cite{Ca90}).
One does not expect to find this effect in the
phenomenological \arma which
knows nothing about the single-particle structure.
And, indeed, there are no rules other than $(K\le I)$
which limit the $K$
projection in the \arma and thus higher $\bar{K}$ values
can be obtained.

At the beginning of this section it was noted that a
deviation from prolate
axial symmetry is a necessary condition for the generation
of $K$-band mixing.
This result is depicted in Fig. \ref{fig:armk} where the
$\gamma$ dependence of
the $\langle ARM | K^2| ARM \rangle_y$ value is shown.
It is therefore interesting to consider changes in the
$\gamma$ deformation
that result when the pairing strength is increased.
The $\gamma$ of a particular \psma wave function can be
determined by taking
advantage of the mapping between the nuclear
deformation variables $(\beta,
\gamma)$ and the $SU(3)$ irrep labels $(\lambda, \,
\mu)$ \cite{LD87,CD88}:
\bea
&&\nn\\
\beta &=& \sqrt{\frac{4\pi}{5}} \frac{1}{ A\bar{r}^2 }
\left( C_2 + 3 \right)^\frac{1}{2}\, ,
\nn\\
&&\label{eq:gam} \\
\gamma &=& \frac{1}{3}
\mbox{ cos}^{-1} \left( \frac{C_3}{2 \left(C_2+
3\right)^\frac{3}{2}} \right)\, ,
\nn
\eea
where $C_2$ and $C_3$ denote
the second and third order Casimir invariants of $SU(3)$
with eigenvalues
\bea
C_2(\lambda, \, \mu) &=&
(\lambda + \mu +3) (\lambda+\mu) - \lambda\mu\, ,  \nn\\
&&\label{eq:c2c3}\\
C_3(\lambda, \, \mu) &=&
  (\lambda - \mu) (\lambda + 2\mu + 3) ( 2\lambda + \mu
+3) \, .
\nn
\eea
Since $C_2$ and $C_3$ are diagonal in the \psma basis
functions
their expectation values are easily calculated.

Using Eqs. \ref{eq:gam}, the dependence of the
$\gamma$ deformation on the
strength of the pairing interaction was determined for the
even $J$ yrast
states of the $(2\pi, 2\nu)$ configuration.
The results are shown in Fig. \ref{fig:psm-gam}.
\\[1 cm]
{\bf Figure \ref{fig:psm-gam}}
\\[1 cm]
With no pairing all members of the yrast band which are
pure $(\lambda,
\, \mu)=(14,0)$ states with the same $\gamma = 3.2^o$
deformation.
But when the strength of the pairing interaction is turned
on, the system
is driven towards triaxiality.
As long as the  spectrum is dominated by the quadrupole-
quadrupole interaction
($G \le 0.1$MeV) and the rotational character of the
eigenstates is more or less
conserved, the yrast eigenstates are less triaxial the higher
their angular momentum.
Since in this domain the change with increasing angular
momentum is a relatively
small effect, it might be attributable to rotational
stretching.
For larger $G$ values, however, the triaxiality is found to
increases more or less
uniformly for all of the yrast states.
For a very large pairing term, where the corresponding
$\gamma$ values are
indicated as ``Asymptotic Values'' on the far right side of
Fig. \ref{fig:psm-gam},
the states show about the same $\gamma$-deformation.
\subsection{Quadrupole Moments and B(E2) Values}
The effect of the pairing interaction on quadrupole
moments and B(E2;$I
\rightarrow I+2$) transition rates is considered in this
subsection.
Recall that increasing the pairing strength corresponds to
increasing the
triaxiality of the calculated yrast states of a system.
To further probe the effect pairing has on collective
behavior it is helpful
to look at the $\gamma$ dependence of quadrupole
moments and B(E2) values.
And as above, it is useful to start with the complementary
\arma results.
\\[1 cm]
{\bf Figure \ref{fig:armq}}
\\[1 cm]
{\bf Figure \ref{fig:psmq}}
\\[1 cm]
Figure \ref{fig:armq} depicts the $\gamma$ dependence
of the spectroscopic
quadrupole moments of even $J$ yrast states of the \arm.
For a range of $\gamma$ values induced by increasing
the pairing strength
from weak to strong in the \psma, that is, for $\gamma$
values between about
$3^o \le \gamma \le 17^o$ (Fig. \ref{fig:psm-gam}), the
negative quadrupole
moments are found to decrease in magnitude for all values
of the angular
momentum.
Comparing this with the corresponding \psma results
shown in Fig. \ref{fig:psmq},
shows that the quadrupole moments of the even yrast
states in the \psma also
decrease with increasing pairing strength.
The similarity of these results confirms that an increase in
the pairing
strength goes hand-in-hand with an increase in the
$\gamma$ deformation.

The situation is less clear if the yrast intra-band B(E2)
values of the two
models are compared, see Figures \ref{fig:armbe2} and
\ref{fig:psmbe2}.
\\[1 cm]
{\bf Figure \ref{fig:armbe2}}
\\[1 cm]
{\bf Figure \ref{fig:psmbe2}}
\\[1 cm]
These figures depict, respectively, the dependence on
$\gamma$ of the
intra-yrast band transitions B(E2;$ I \rightarrow I+2$) in
the asymmetric
rotor and pseudo-SU(3) models.
While for $3^o \le \gamma \le 17^o$ the \arma results
show a moderate decrease
in the B(E2;$ 0 \rightarrow 2$) strength, for all pratical
purposes the \arma
B(E2;$ I \rightarrow I+2$) values are independent of
$\gamma$.
In contrast with this, Fig. \ref{fig:psmbe2} shows that
the \psma yields a very
pronounced drop in the  B(E2;$ I \rightarrow I+2$)
transition strengths with
increasing pairing strength.
So while diagonal measures suggest that increasing
$\gamma$ in the \arma and
the pairing strength in the \psma produce similar effects,
they predict quite
different results for intra-yrast band B(E2) transitions.

\section{Application to $^{140}_{58}$Ce$_{82}$
\label{se:fit}}

The results of the previous section form a backdrop for
\psma applications
that describe and predict nuclear properties  like excitation
energies, B(E2) values,
quadrupole moments, and $g_R$-factors.
A more complete and systematic study of rare-earth nuclei
will be reported elsewhere
\cite{TE95}; this section focuses on the semi-magic
$^{140}_{58}$Ce$_{82}$ nucleus
which has eight valence protons outside the $Z = 50$
core.
This means the model space does not have to be
truncated, nevertheless, the
$^{140}_{58}$Ce$_{82}$ system is sufficiently well-
studied experimentally to
allow for a thorough examination of the \psma description
of its properties.

The $^{140}_{58}$Ce$_{82}$ nucleus has six valence
protons in the usual $(N_\pi=4)$
shell, or equivalently, six protons in the pseudo
$(\tilde{N}_\pi=3)$ space, and no
valence neutrons.
The neutron part of the total wave function is therefore
trivial and this in turn
implies some simplifications in the Hamiltonian of Eq.
\ref{eq:psh}.
Dropping all terms with vanishing matrix elements yields
\beq
H_{PSU(3)} =
-\frac{\chi}{2} \,\,  Q^a \cdot Q^a
- G_\pi H^\pi_P
+ a K_J^2 + bJ^2
+ D_\pi \sum_{i_\pi}l_{i_\pi}^2
+ c \,C_3
\label{eq:ceham}
\eeq
where the third order Casimir operator, $C_3$, of
$SU(3)$ has been added.
The physical meaning of this operator, see Eqs.
\ref{eq:gam} and \ref{eq:c2c3},
can be easily understood through the relation of the \psma
to the geometric
collective model.
As mentioned in the introduction, this liquid-drop type
model describes the
low-energy collective excitations of even-even nuclei in
terms of quadrupole
surface vibrations and rotations of the nucleus as a whole.

A number of publications show that a phenomenological
treatment of the geometric
collective model provides a very successful means for
describing the low-energy
properties (energies, B(E2) values, and quadrupole
moments) of even-even nuclei
throughout, including phenomena like shape coexistence
and shape transition
\cite{TM91,EG87}.
The basis of such an approach is an expansion of the
model Hamiltonian in powers
of the collective $(\beta, \gamma)$ variables.
A typical ansatz for the nuclear potential in such a scheme
is
\beq
V(\beta, \, \gamma) = x_2 \beta^2 + x_3 \beta^3 \cos
3\gamma + x_4 \beta^4 + \,
\ldots \label{eq:V}
\eeq
where the real numbers $x_2, x_3, x_4, \, \ldots$ are
usually determined in a
least-square fitting procedure aimed at an optimal
description of the experimental
data of the nucleus under consideration.

To understand the physical effect of $C_3$ in Eq.
\ref{eq:ceham}, it is useful
to recall the mapping between the pseudo-SU(3) labels
$(\lambda, \, \mu)$ and
the deformation variables ($\beta, \, \gamma)$
\cite{LD87}  that is given in
Eq. \ref{eq:gam}.
The mapping is based on an associated of the invariants
of the $SU(3)$ and
geometric collective models,
\beq
C_2 \sim \beta^2, \, \, \, \, \, \, \, \,  C_3 \sim \beta^3 \cos
3\gamma. \,
\eeq
These relations also provide an explanation for the
physical meaning of $C_3$ in
Eq.\ref{eq:ceham}:
Depending on the positive or negative sign of $c$ in Eq.
\ref{eq:ceham}, or
equivalently on the sign of $x_3$ in Eq. \ref{eq:V}, the
nuclear system is
driven towards more oblate or more prolate deformations.
So in contrast with the quadrupole-quadrupole
interaction, the $C_3$ term adresses
the $\gamma$ degree-of-freedom and helps determine the
triaxiality of the system.

Six parameters (see Table \ref{tab:Cepara}) are required
to fix the \psma Hamiltonian
(Eq. \ref{eq:ceham}) for $^{140}_{58}$Ce$_{82}$.
These values were determined in a best-fit calculation
based on known experimental
energies and B(E2) values of the system.
\\[1 cm]
{\bf Table \ref{tab:Cepara}}
\\[1 cm]
There are a number of theoretical and experimental studies
which focused on the
$^{140}$Ce nucleus \cite{Ce all}.
The current study focuses only on its excitation spectrum;
the reader is referred
to \cite{TE95} for other measures as well as a systematic
comparison of the properties
of several different rare earth nuclei.
\\[1 cm]
{\bf Figure \ref{fig:Ceen}}
\\[1 cm]
Pseudo-SU(3) model results for $^{140}$Ce energies up
to about 3 MeV are shown in
Fig. \ref{fig:Ceen}.
A comparision with experimental excitation energies that
was taken from \cite{NDS}
is given to the immediate right of the pseudo-SU(3)
results followed by the spectrum
obtained by Wildenthal \cite{Wi69} and the results of a
model study by Waroquier and
Heyde \cite{WH71}.

The Wildenthal results were obtained in a conventional
shell  model calculation that
used a modified surface $\delta$ interaction in a truncated
configuration space.
The single-particle energies and the parameters of the
modified surface $\delta$
interaction were adjusted to describe the  excitation
energies of nuclei in the mass
range from $A=136$ to $A=145$ \cite{Wi69}.
In addition to a surface $\delta$ interaction, Waroquier
and Heyde \cite{WH71} used
a Gaussian residual nucleon-nucleon interaction.
The single-particle energies were determined by solving
the inverse-gap equations.

While the agreement between the three theories is of
approximately the same quality
for energies below about 2.5 MeV, the \psma does not
account for a low-lying $5^+$
state which experimental evidence suggests lies at 2.349
MeV.
Otherwise, both the \psma and the calculations by
Wildenthal account for virtually
all levels up to about 3 MeV while the results by
Waroquier and Heyde seem to miss
a few of the experimentally established states.
In addition, note that the two former models predict a few
levels, like two $4^+$
states at about 2.8 MeV, which have not been found
experimentally.
The pseudo-SU(3) spectrum also predicts a $0^+$ state at
about 2.25 MeV that cannot
be found in the experimental results.

\section{Summary, Conclusion and Outlook}
The effect of pairing on $K$-band mixing, B(E2) values,
and quadrupole moments
has been studied within the framework of a complete
(untruncated) \psma theory.
The \arm, which is a collective model theory that
describes nuclear properties
by means of rotations only, was introduced as a backdrop
for helping to identify
collective properties of the \psma and, in a complementary
way, for discovering
limitations of the collective model approach:
\begin{itemize}
\item From general considerations it is clear that
deviations from axial symmetry
      are a necessary condition for $K$-band mixing.
      This was seen explicitly for the \arma by noting that
for $\gamma \ne 0$
      the non-diagonal part of the Hamiltonian is non-zero,
and for the \psma since
      an increase in the pairing strength not only induces
$K$-band mixing but also
      generates an increase in $\gamma$ deformation.
\item The $K$ label of the \psma is limited by a $SU(3)$
to $SO(3)$ selection rule.
      This restriction is  inherent to the Pauli Exclusion
principle and not part
      of the \arma where, in principle, there is no maximum
$K$ value.
\item The quadrupole moments of yrast states of the \arma
decrease in magnitude
      with increasing $\gamma$ deformation ($0 \le
\gamma \le 30^o$) and in the
      \psma with increasing pairing strength.
      Since an increase in the pairing strength was found to
be correlated with an
      increase in $\gamma$ deformation from about
$\gamma=3^o$ to $\gamma=17^o$,
      the physics behind the two phenomena appears to be
very similar.
\item The yrast intra-band B(E2;$ I \rightarrow I+2$)
strengths decrease rather
      strongly with increasing pairing strength while the
\arma values show only
      a moderate decrease with increasing $\gamma$
deformation.
\end{itemize}
Finally, the \psma was used for a description and
prediction of experimental
energies of $^{140}$Ce and the results were compared to
other theories.
The Hamiltonian parameters were determined in a best-fit
calculation that used
as input the experimental energies and the B(E2)-values
of the low-lying states.
The \psma was found to describe the known experimental
energies satisfactorily,
in a addition, a few new level were predicted.
A more complete and systematic comparison of the
predictions of the pseudo-SU(3)
model with experimental data and other theories  will be
reported in a forthcoming
contribution \cite{TE95}.

Regarding future developments it should be obvious that
the presented formalism
can be easily applied to $(S \ne 0)$ states and hence to a
description of not only
even-even but also odd-even, even-odd, and odd-odd
nuclei.
To put this study into perspective, it is also important to
note that the results
learned from \psma studies apply as well to the pseudo-
symplectic model which is a
natural extension that takes couplings to higher shell fully
into account.

\newpage

\begin{thebibliography}{10}

\bibitem{AH69}
{A. Arima, M. Harvey and K. Shimizu, Phys. Lett. {\bf
30B} (1969) 517}.

\bibitem{HA69}
{K.T. Hecht and A. Adler, Nucl. Phys. {\bf A137}
(1969) 129}.

\bibitem{TD94}
{D.Troltenier, J.P.Draayer, P.O. Hess, O. Castanos,
Nucl. Phys.,
{\bf 576} (1994) 351}.

\bibitem{RR73}
{R.D. Ratna-Raju, J.P. Draayer and K.T. Hecht, Nucl.
Phys. {\bf A202}
(1973) 433}.

\bibitem{DW84}
{J.P. Draayer and K.J. Weeks, Ann. Phys. {\bf 156 }
(1984) 41}.

\bibitem{CD87}
{O. Casta\~{n}os, J.P. Draayer and Y. Leschber, Ann.
Phys. {\bf 180}
(1987) 290}.

\bibitem{NT90}
{W. Nazarewicz, P.J. Twin, P. Fallon, J.D. Garrett,
Phys. Rev. Lett. {\bf C64}
(1990) 1654}.

\bibitem{CH94}
{O. Castanos, J.G. Hirsch, O. Civitarese and P.O. Hess,
Nucl. Phys.
{\bf A571} (1994) 276}.

\bibitem{BD94}
{C. Bahri and J.P. Draayer, Comp. Phys. Comm. {\bf
83} (1994) 59}.

\bibitem{TD94a}
{D. Troltenier, C.Bahri, and J.P. Draayer, accepted by
Nucl. Phys. A (1994)}.

\bibitem{TM91}
{D. Troltenier, J. Maruhn, and P.O. Hess, in:
{\em Computational Nuclear Physics 1} ed. by K.
Langanke, J. Maruhn, S.E. Koonin
(1991)}.

\bibitem{AD73}
{Y. Akiyama and J.P. Draayer, Comp. Phys. Comm.
{\bf 5} (1973) 405}.

\bibitem{ND90}
{H.A. Naqvi and J.P. Draayer, Nucl. Phys.  {\bf A516}
(1990) 351}.

\bibitem{ND92}
{H.A. Naqvi and J.P. Draayer, Nucl. Phys.  {\bf A536}
(1992) 297}.

\bibitem{TE95}
{D. Troltenier, J. Escher and J.P. Draayer,
{\it Rare earth nuclei in the psedo-SU(3) model}
submitted to Nucl. Phys.}

\bibitem{VM88}
{D.A. Varshalovich, A.N. Moskalev, and V.K.
Khersonskii, in:
{\em Quantum Theory of Angular Momentum} (World
Scientific, Singapore, 1988)}.

\bibitem{DF58}
{A.S. Davidov, B.F. Fillipov, Nucl. Phys.
{\bf 8} (1958) 237}.

\bibitem{DC60}
{A.S. Davidov, A.A. Chaban, Nucl. Phys.
{\bf 20} (1960) 499}.

\bibitem{Da65}
{J.P. Davidson, Rev. Mod. Physics {\bf 37} (1965)
105}.

\bibitem{AB68}
{S.M. Abecasis, H.E. Bosch, A. Plastino, Nuovo
Cimento {\bf 54B} (1968) 245;
Nucl. Phys {\bf A129} (1969) 434}.

\bibitem{EG87}
{J. Eisenberg and W. Greiner, in:  {\em Nuclear Theory,
Vol. 1, Nuclear Models}
(North-Holland Physics Publishing, 1987)}.

\bibitem{Ro57}
{M.E. Rose, in:  {\em Elementary Theory of Angular
Momentum}
(Wiley, New York, 1957)}.

\bibitem{BJ91}
{H.G. B\"orner, J. Jolie, S.J. Robinson, B. Krusche, R.
Piepenbring, R.F. Casten,
A. Aprahamian, J.P. Draayer, Phys. Rev. Lett. {\bf 66}
(1991) 691}.

\bibitem{WA94}
{X. Wu, A. Aprahamian, S.M. Fischer, W.Reviol,
G.Liu, J.X. Saladin,
Phys. Rev. {\bf C49} (1994) 1837}.

\bibitem{TM92}
{D. Troltenier, J. Maruhn, W. Greiner and P.O. Hess,
Z.Phys.{\bf A343} (1992) 25}.

\bibitem{NB94}
{H.A. Naqvi, C. Bahri, D. Troltenier, J.P. Draayer and
A. Faessler,
accepted  by Z. Phys. A - Hadrons and Nuclei (1994)}.

\bibitem{CL93}
{R.F. Casten, P.O. Lipas, D.D. Warner, T. Otsuka, K.
Heyde and
J.P. Draayer, in:  {\em Algebraic Approaches to Nuclear
Struture},
ed. by R.F. Casten (Harwood Academic Publishers,
1993)}.

\bibitem{Ca90}
{R.F. Casten in:  {\em Nuclear Structure from a Simple
Perspective}
(Oxford University Press, 1990)}.

\bibitem{LD87}
{Y. Leschber, J.P. Draayer,  Phys. Lett. {\bf B190}
(1987) 1}.

\bibitem{CD88}
{O. Casta\~{n}os, J.P. Draayer and Y. Leschber,
Z.Phys.{\bf A329} (1988) 33}.

\bibitem{Ce all}
{E. Michelakakis and W.D. Hamilton,
J.Phys. {\bf G8} (1982) 581.

I. Dioszegi, A. Veres, W. Enghardt, and H. Prade,
J. Phys. {\bf 11} (1985) 853;

W. Enghardt, L. K\"aubler, H.Prade, H.-J. Keller, and
F. Stary,
Nucl. Phys. {\bf A449} (1986) 417;

F. Monti, G. Bonsignori, M. Savoia, and Y.K.
Gambhir,
Nuovo. Cim. {\bf 104 A} (1991) 33;

D. Bazzacco, F. Brandolini, K. L\"owenich, P. Pavan,
C. Rossi-Alvarez, E. Maglione,
M.de Poli, and A.M.I.Haque,
Nucl.Phys. {\bf A533} (1991) 541;

M. Grinberg, Thai Khac Dinh, C. Protochristov, I.
Penev, C. Stoyanov, and
W. Andrejtscheff,
J. Phys. {\bf G19} (1993) 140.}

\bibitem{NDS}
{Nuclear Data Sheets, ed. by National Nuclear Data
Center,
Brookhaven National Laboratory, Upton, NY 11973,
USA.
Updated experimental data were obtained via e-mail at
BNLND2.DNE.BNL.GOV}

\bibitem{Wi69}
{B.H. Wildenthal,
Phys. Rev. {\bf C22} (1969) 1118.}

\bibitem{WH71}
{M. Waroquier and K. Heyde,
Nucl.Phys. {\bf A164} (1971) 113.}

\end{thebibliography}

\newpage
values
\begin{table}
\begin{center}
\begin{tabular}{|r|c|}
\hline
 &\\
Parameter &  $^{140}$Ce \\
 & \\
\hline
 &\\
$\chi$ [keV]    &  3.97   \\
$ D_\pi $ [keV] & -101.05 \\
$ G_\pi $ [keV] &  157.5  \\
$ a     $ [keV] &  130.6  \\
$ b     $ [keV] &  0.085  \\
$ c     $ [keV] &  5.078  \\
& \\
\hline
\end{tabular}
\end{center}
\caption{List of parameters used in the pseudo-SU(3)
calculations of $^{140}$Ce
(see text).
\label{tab:Cepara}}
\end{table}

\newpage
\clearpage
\noindent
\begin{center}
{\Large Figure captions}
\end{center}
\noindent
\begin{figure}[h]
\caption{\label{fig:armen}
The $\gamma$ dependence of the lowest \arma eigen-
energies in units of
$\frac{\hbar^2}{4B \beta^2}$.
The angular momentum values are indicated on the far
right and along the central
vertical line.
}
\end{figure}
\noindent
\begin{figure}[h]
\caption{\label{fig:armk}
The $\gamma$ dependence of the expectation value
$\langle ARM |\hat{K}^2| ARM \rangle_y$
of the $\hat{K}^2$ operator is shown for even angular
momentum yrast states of the \arm.
The angular momentum of the various curves are
indicated on the far right.
The values for $\langle ARM |\hat{K}^2| ARM
\rangle_y$ that are obtained for complete
mixing, that is, when all the diagonalization coefficients
are set equal to one
another, are indicated by horizontal lines at the far right
labeled above with the
corresponding value of the angular momentum.
}
\end{figure}
\noindent
\begin{figure}
\caption{\label{fig:2p2nk**2}
The expectation value of the $\hat{K}^2$ operator in
even angular momentum yrast states
is shown as a function of the pairing strength parameter
$G$, where for simplicity the
latter were all set equal, $G_\pi = G_\nu = G$.
The angular momentum value of each curve is indicated
on the far right.
The horizontal lines on the far right indicate asymptotic
values for the matrix
elements $\langle ARM |\hat{K}^2| ARM \rangle_y$ that
are obtained in the limit of a
very large pairing strength.
}
\end{figure}
\noindent
\begin{figure}[h]
\caption{\label{fig:k-av}
This figure illustrates an average property of the \psma
basis states, namely,
the average $K$ value for all basis states of a fixed even
angular momentum $J$.
The $\bar{K}$ distribution shown is for the $(m_\pi=2,
\, m_\nu=2)$ configuration.
}
\end{figure}

\noindent
\begin{figure}[h]
\caption{\label{fig:k-av gen}
This figure is similar to Figure 4.
The $\bar{K}$ values were calculated for all even angular
momentum states $J$ and
different particle numbers $m$ in the $N=3$ shell [ 5a),
left] and the
$N=4$ shell [ 5b), right].
(Particle distributions corresponding to more than half-
filled shells
can be
obtained by
invoking particle-hole symmetry.)
}
\end{figure}
\noindent
\begin{figure}[h]
\caption{\label{fig:psm-gam}
The expectation value of the $\gamma$ deformation,
$\langle \gamma \rangle$,
is shown for even angular momentum yrast states of the
$(m_\pi=2, \, m_\nu=2)$
configuration.
The abscissa is the pairing strength parameter $G$ where
for simplicity the
proton and neutron strengths were set equal, $G_\pi =
G_\nu \equiv G$.
On the far right the $\langle \gamma \rangle$ values for a
very large value of
the pairing strength (`` Asymptotic Values'') are
indicated as bars, each of them
labeled by
its angular momentum value.
}
\end{figure}
\noindent
\begin{figure}[h]
\caption{\label{fig:armq}
Quadrupole moments of yrast states of the \arma for even
values of the angular
momentum are shown as a function of the $\gamma$
deformation where the numbers
on the far left and right denote the angular momentum in
units of $\hbar$.
}
\end{figure}
\noindent
\begin{figure}[h]
\caption{\label{fig:psmq}
Quadrupole moments of yrast states of the $(m_\pi=2, \,
m_\nu=2)$ configuration
for even values of the angular momentum are shown as a
function of the pairing
strength $G$.
The numbers next to the curves on the
right denote the value of the angular
momentum in units of $\hbar$.
For comparison, values for the prolate rotor are indicated
on the far left.
Note that these are almost identical to the \psma values.
On the far right quadrupole moments for the case of very
large $G$ are indicated
by dashed horizontal lines (`` Asymptotic Values '').
}
\end{figure}
\noindent
\begin{figure}[h]
\caption{\label{fig:armbe2}
The $\gamma$ dependence of the intra-band transition
probabilities,
B(E2;$I \rightarrow I+2)$, for the \arma yrast states are
shown, with the initial
and final values of the angular momentum indicated on
the far right.
}
\end{figure}
\noindent
\begin{figure}[h]
\caption{\label{fig:psmbe2}
The intra-band reduced transition probabilities, B(E2;$I
\rightarrow I+2)$, for
eigenfunctions of the \psma are shown as a function of
increasing pairing strength.
On the far right the horizontal lines labeled ``Asymptotic
Values'' are for a very
large value of the pairing strength.
The dotted lines from the smaller $G$ results are included
to help guide the eye.
Similarly, on the far left the horizontal lines labeled
``Rotor Values'' are  the
B(E2;$I \rightarrow I+2)$ strengths of the prolate rotor.
}
\end{figure}
\noindent
\begin{figure}[h]
\caption{
\label{fig:Xeen}
 From left to right the figure depicts the
excitation spectrum of $^{140}$Ce as calculated within
the \psma, the experimental
values (left center), the results of a calculation by
Wildenthal (right center),
and, on the very right, the energies as obtained in the
model used by Waroquier and
Heyde (see text).
Note that the thin dashed lines between the levels are
supplied to guide the eye
between corresponding levels and that the spacings
between the ground and the first
excited states are not to scale.)
\label{fig:Ceen}
}
\end{figure}
\end{document}